\documentclass[
 reprint,
 amsmath,amssymb,
 aps,
]{revtex4-2}

\usepackage{graphicx}
\usepackage{dcolumn}
\usepackage{bm}
\usepackage{xcolor}

\begin{document}

\definecolor{ultramarine}{rgb}{0.07, 0.04, 0.56}
\newcommand{\is}[1]{\textcolor{ultramarine}{\textbf{(IS: #1)}}}

\newcommand{\apjs}{ApJS}
\newcommand{\apjl}{ApJ Letters}
\newcommand{\mnras}{MNRAS}
\newcommand{\aap}{Astronomy \& Astrophysics}
\newcommand{\jcap}{Journal of Cosmology and Astro-Particle Physics}

\preprint{APS/123-QED}

\title{The Galactic Center $\gamma$-ray Excess and the Look-Elsewhere Effect}

\author{Isabel S. Sands}
\affiliation{Kavli Institute for Astrophysics and Space Research, Massachusetts Institute of Technology, Cambridge, MA, 02139, USA}
\author{Philip F. Hopkins}%
\affiliation{TAPIR, California Institute of Technology, Mailcode 350-17, Pasadena, CA 91125, USA}

\date{\today}

\begin{abstract}
The origin of the excess of GeV $\gamma$-rays in the Milky Way's galactic center observed by \textit{Fermi}-LAT has been an open question for nearly two decades. The two leading interpretations of this signal are that it is produced either by annihilating dark matter (DM), or arises from unresolved point sources, such as a novel population of millisecond pulsars. In both scenarios, uncertainties and limitations in the modeling of galactic diffuse $\gamma$-ray emission, such as the steady-state approximation of cosmic ray (CR) propagation and assumption of an axisymmetric distribution of CRs, complicate the interpretation of the galactic center excess (GCE).  Simulations with full magnetohydrodynamics (MHD) and CR transport can provide a new avenue for understanding the astrophysical phenomena that drive galactic diffuse $\gamma$-ray emission. We utilize MHD simulations of Milky Way analogs with time-dependent, spectrally-resolved CR transport to reconstruct diffuse $\gamma$-ray emission. Comparing instantaneous emission to a time-averaged baseline, we find residuals up to several times larger than the Fermi GCE, with a wide range of spectral shapes and spatial morphologies. At some simulation snapshots, the spectrum and morphology of the residuals resemble the Fermi GCE. Our results suggest that the Fermi GCE may be a manifestation of the look-elsewhere effect — a transient fluctuation in diffuse emission from stellar sources, observed at an inopportune moment — and motivate the development of new Fermi-LAT analysis templates that account for time-dependent CR transport.
\end{abstract}

\maketitle



\section{Introduction}
Observations from the Fermi Large Area Telescope (LAT) have detected an excess of GeV gamma-rays within the central $\sim 10^{\circ}$ of the Milky Way relative to known astrophysical sources and models of diffuse $\gamma$-ray emission \cite{2011PhLB..697..412H, 2011PhRvD..84l3005H, 2017ApJ...840...43A}. More than a decade later, the origin of this signal remains debated, with some literature arguing that the observed excess is a smoking gun for DM annihilation \cite{2016PDU....12....1D,2019arXiv190408430L,Hooper:2022bec}, and other studies attributing it to other phenomena such as unresolved point sources (e.g., millisecond pulsars) \cite{Lee_2016, Bartels_2016}. Both of these posited explanations have faced challenges: in the case of the annihilating DM hypothesis, some studies have found that the morphology of the excess traces the galactic bulge, indicating an astrophysical origin \cite{Macias_2018, 2018NatAs...2..819B, Song2024}. However, the morphology of the galactic center (GC) $\gamma$-ray excess is a matter of ongoing study and debate, with no conclusive answer \cite{2020PhRvD.102f3019L, 2020PhRvL.125l1105L,2023MNRAS.522L..21M, Ramirez2025}. The pulsar scenario presents its own set of challenges: the population of progenitor x-ray binaries in the galactic center is not sufficient to produce the number of millisecond pulsars that the $\gamma$-ray excess requires \cite{2017JCAP...05..056H}, although the requisite pulsar population could still be generated through more exotic channels \cite{Gautam2022}. Additionally, recent work has demonstrated that the methods used in some past analyses that favored the pulsar scenario can falsely interpret an injected, spatially-smooth signal consistent with annihilating DM as a collection of point sources \cite{2020PhRvL.125l1105L, 2020PhRvD.102f3019L}.

While most of the debate surrounding the GCE has focused on DM and millisecond pulsar origins of the signal, other astrophysical sources have been considered, including non-thermal Bremsstrahlung produced in cosmic ray interactions with molecular clouds near the GC \cite{Macias2014, Macias2014a}, and past activity in the GC from either an AGN or supernovae \cite{Petrovic2014, Macias_2018}. However, other studies have suggested that the predominantly leptonic bursts in these scenarios are inconsistent with the uniform spectral shape of the GCE, unless the CR parameters are finely tuned \cite{Cholis2015, 2022PhRvD.105j3023C}. A major limitation of nearly all past studies of the galactic center excess (GCE) is that they rely on steady-state CR propagation codes that make simplistic assumptions, such as axisymmetric distribution of CRs, and cannot model time-dependent phenomena such as galactic winds, gas coupling to magnetic fields, or turbulence.

Recent numerical developments that enable the simulation of spectrally-resolved CRs in MHD galaxy formation simulations can provide a complementary approach to modeling cosmic ray transport and propagation in galactic environments. It has been demonstrated that CRMHD simulations of Milky Way-like galaxies from cosmological initial conditions can reproduce local interstellar medium (ISM) observations of the CR proton and electron spectra, positron fraction, and $\gamma$-ray emissivity \cite{Hopkins_CR, Sands2025}. In particular, \cite{Sands2025} utilized such simulations to model diffuse $\gamma$-ray emission in the galactic centers of Milky Way-like galaxies from neutral pion ($\Pi^0$) decay, non-thermal relativistic Bremsstrahlung, and inverse Compton scattering of CR leptons, finding that the highly dynamical environment drives order-of-magnitude fluctuations in the $\gamma$-ray spectrum on Myr timescales from supernova-sourced CRs alone. 

With the ability to simulate diffuse $\gamma$-ray emission in full astrophysical context, it is worthwhile to revisit the extent to which CR physics can contribute to the Fermi signal without invoking additional sources. Additionally, simulations allow us to study multiple realizations of Milky Way-mass halos at multiple snapshots in time. In analyzing these simulations, we find that there are transient bumps and wiggles in the GC $\gamma$-ray spectrum, accompanied by highly-variable morphology of diffuse emission from CR physics. Thus, in this letter, we consider the possibility that the Fermi-LAT galactic center $\gamma$-ray excess is a manifestation of the look-elsewhere effect-- have we caught the Milky Way's GC at an (in)opportune time? Furthermore, are there signatures that can distinguish this scenario from other explanations of the galactic center excess?


This letter is structured as follows. We first provide a brief description of the simulations and the methods used to model $\gamma$-ray emission from the simulated CRs, and define the baseline ``templates" used to calculate residuals at given snapshots of the simulations. We then present the range of residual spectra and morphologies that occur due to time-dependent CR transport. Finally, we show how transient behavior in the GC diffuse $\gamma$-ray emission can mimic the measured Fermi-LAT galactic center excess, and discuss the implications of this result.

\section{Methods}

\label{sec:methods}

 We analyze a set of three simulations of Milky Way (MW)-like galaxies (labeled \textbf{m12f}, \textbf{m12i}, and \textbf{m12m}), first introduced in \cite{Hopkins_CR}, run with FIRE-2 physics and multi-species, spectrally-resolved cosmic rays. These galaxies are simulated from cosmological initial conditions at $z\sim 100$, and evolved to late times with the single-bin fluid CR treatment described in \cite{Chan_2019, Hopkins_2019}; at redshift $z = 0.05$, the simulations are reinitialized with CR spectra for protons, electrons, anti-protons, positrons, and heavier nuclei using spectral shapes and relative normalizations from local ISM observations. CR are produced by supernovae, with 10\% of pre-ejecta kinetic energy going into the CRs, and 2\% of that energy into the leptons. These simulated galaxies have been widely studied by members of the FIRE collaboration, and have been demonstrated to be good MW analogs in a variety of metrics. However, we ultimately do not consider \textbf{m12i} in this work due to this specific realization featuring a significant warp in the disk that is not compatible with the MW's morphology (see \cite{Sands2025}). The two remaining galaxies, \textbf{m12f} and \textbf{m12m}, are both Milky Way-mass spiral galaxies; \textbf{m12f} has a flatter, extended disk, and \textbf{m12m} has a more pronounced galactic bulge. 

\begin{figure}
    \centering
    \includegraphics[width=\linewidth]{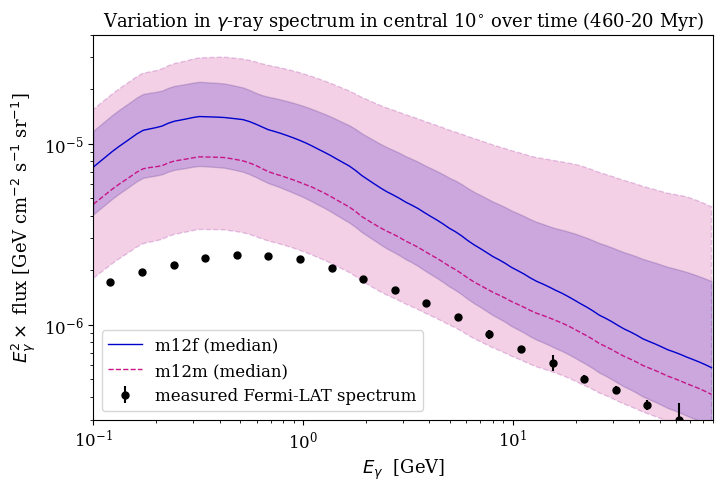}
    \caption{The $\gamma$-ray spectrum in the central $10^{\circ}$ region of interest (ROI) for simulated galaxies \textbf{m12f} and \textbf{m12m}; the median value of the spectrum across 460-20 Myr in lookback time is shown in the solid and dashed lines, respectively. The shaded regions indicate the full range of spectra over the same time span. The measured Fermi-LAT $\gamma$-ray spectrum in the same ROI of the Milky Way is shown by the black dotted line. While features of the $\gamma$-ray spectra for the simulated galaxies show similarities to the measured spectrum, such as a peak below 1 GeV and a similar slope of the high-energy tail, the simulated spectra are an order-of-magnitude larger, due to the greater number of stellar sources of CRs in the simulations compared to the Milky Way. These features are discussed in detail in \cite{Sands2025}. }
    \label{fig:GC_spectrum}
\end{figure}

We model diffuse $\gamma$-ray emission from $\Pi^0$ decay, non-thermal relativistic Bremsstrahlung, and inverse Compton scattering in the simulations according to the procedure described in \cite{Sands2025}. We note that both \textbf{m12f} and \textbf{m12m} have higher star formation rates (SFRs) within the central $\sim$ kpc than that observed in the MW, resulting in a factor of a few to 10 times more diffuse $\gamma$-ray flux than that observed by \textit{Fermi}-LAT (Figure \ref{fig:GC_spectrum}). However, the star formation rates in both galaxies are highly variable over time; past analysis of simulations with the same initial conditions but different FIRE physics showed that the SFR of \textbf{m12f} is consistent with that of the Milky Way's Central Molecular Zone at some snapshots \cite{Orr2021}. It is likely that including active galactic nuclei (AGN) feedback would suppress the SFR in the GC, but for the purpose of isolating effects due to different sources of CRs, we do not include AGN in these simulations. We do not model $\gamma$-ray emission from either DM annihilation or decay, or from a millisecond pulsar population; the focus of this work is the time-variability of diffuse $\gamma$-ray from known stellar sources. 

\begin{figure*}
    \centering
    \includegraphics[width=0.8\linewidth]{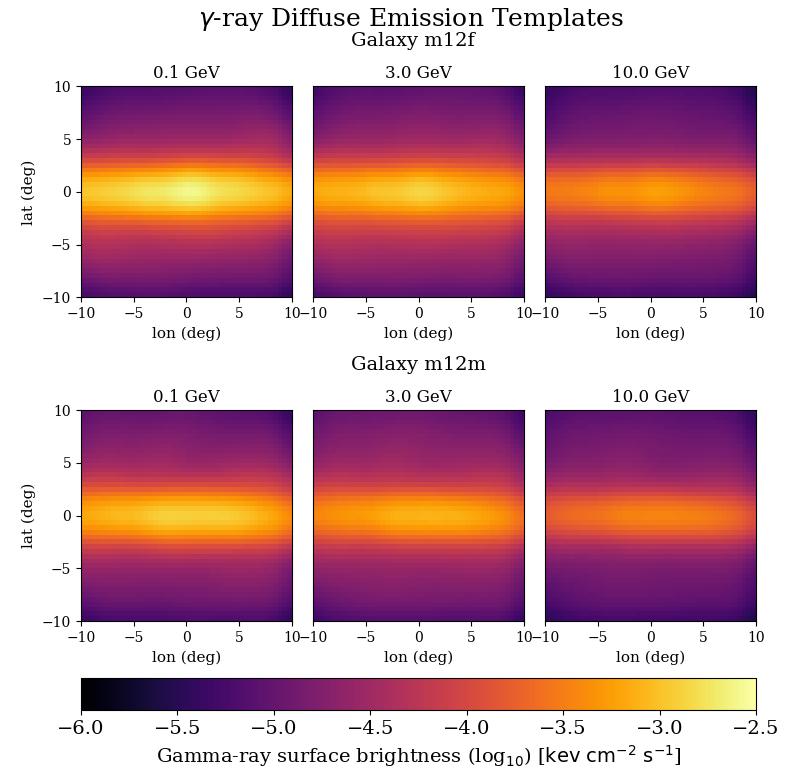}
    \caption{The baseline diffuse $\gamma$-ray emission template for galaxy \textbf{m12f} (top), and galaxy \textbf{m12m} (bottom) shown in three different energy channels (0.1 GeV, 3 GeV, and 10 GeV). The template is constructed by taking the mean value of flux for snapshots in which the total $\gamma$-ray flux in the ROI is less than the median across all snapshots for that simulation. Note that the template for \textbf{m12f} is slightly brighter than the template for \textbf{m12m}, although both templates have comparable morphologies and disk thicknesses.}
    \label{fig:template}
\end{figure*}

The principal exercise of this work is to compare the $\gamma$-ray emission at each simulation snapshot to a baseline ``template" that approximates steady-state $\gamma$-ray emission in the galactic center.   We construct the baseline template as follows: we calculate the total $\gamma$-ray flux in the region of interest (ROI) for each simulation snapshot. A snapshot is defined as having ``low-burstiness" if the total flux at that snapshot is below the median value across all snapshots for that simulation. The baseline template is the mean of the low-burstiness snapshots (Figure \ref{fig:template}). The ``residuals" considered are the difference between the flux at a given snapshot and the baseline template.

At each snapshot, we compare the diffuse $\gamma$-ray spectrum and morphology in the ROI to that of the baseline template, considering both 10$^\circ$ and 20$^\circ$ ROIs. The center of the galaxy is defined as the mass-weighted centroid of star particles in the simulation. Additionally, to determine the effect of diffuse emission along the line-of-sight, we consider three different radial cuts on the data ($r_{\rm{cut}}$ = 0, 2, and 7 kpc), with $r_{\rm{cut}}$ defined as distance from the observer.



\section{Results}

\subsection{Overview}

\begin{figure*}
    \centering
    \begin{tabular}{c c }
        \includegraphics[width=0.51\linewidth]{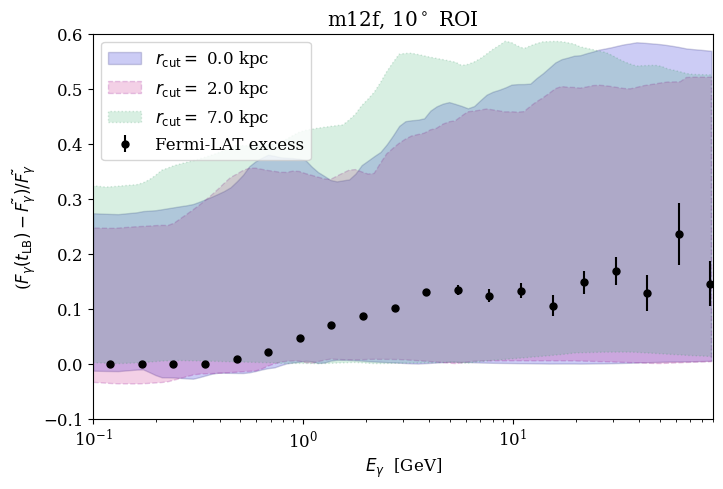}& \includegraphics[width=0.49\linewidth]{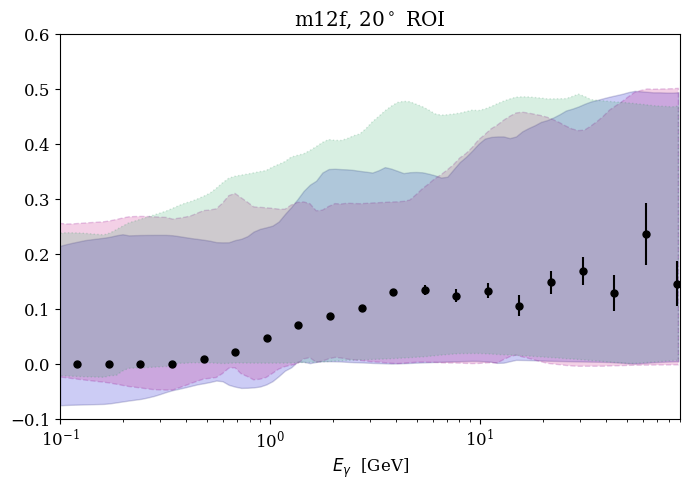}\\
         \includegraphics[width=0.51\linewidth]{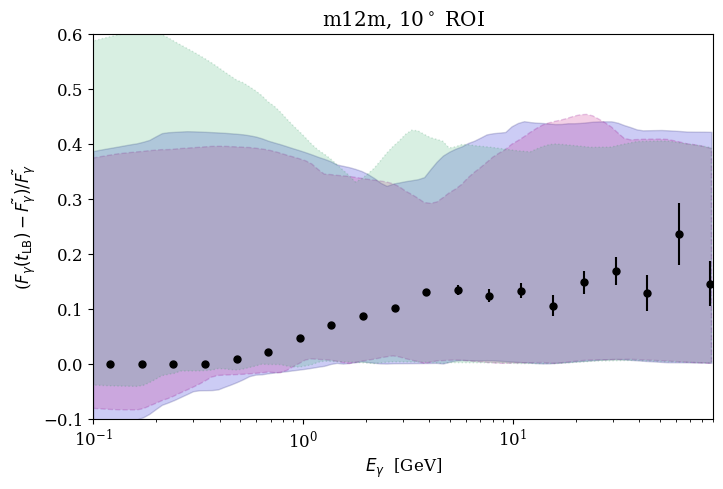}& \includegraphics[width=0.49\linewidth]{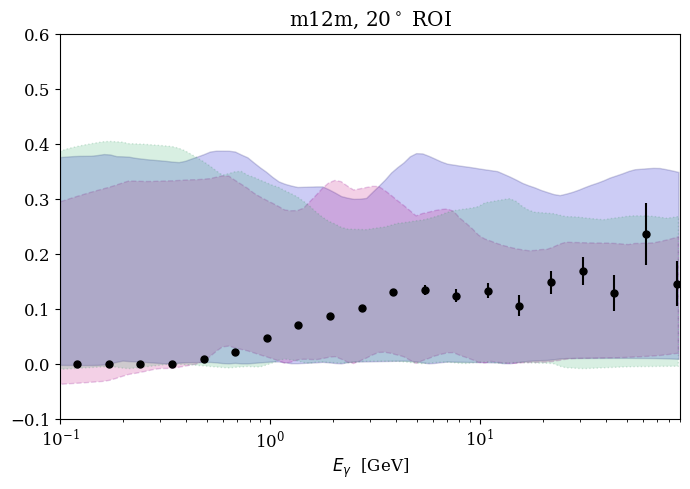}\\
     \end{tabular}
    \caption{The range of spectra of residuals for galaxies \textbf{m12f} (top) and \textbf{m12m} (bottom), in $10^{\circ}$ (left) and $20^{\circ}$ (right) ROIs; the shaded region shows the 25th and 75th percentiles from snapshots across a 460 to 20 Myr span. Different radial cuts ($r_{\rm{cut}} =$ 0, 2, and 7 kpc) are shown by the different colored regions (blue, pink, and green, respectively). The measured Fermi-LAT excess in the $10^{\circ}$ ROI is shown by the black dotted line. For all combinations of ROIs and radial cuts considered, the measured Fermi-LAT excess falls comfortably within the 25th and 75th percentiles, with many residuals considerably larger in magnitude.}
    \label{fig:spec_spread}
    \index{figures}
\end{figure*}

In this section, we consider both the spectral shape and morphology of the residual $\gamma$-ray flux relative to the baseline template. For both galaxies \textbf{m12f} and \textbf{m12m}, the spectrum and morphology of the residuals is highly variable over time, for all configurations of ROI and radial cuts considered. Excesses relative to the baseline template are commonly much larger in magnitude than the measured Fermi-LAT excess (Figure \ref{fig:spec_spread}). In general, the 10$^{\circ}$ ROI shows more variability over time than the larger 20$^{\circ}$ ROI; this is due to the inclusion of regions of the galactic disk and the Fermi bubble-like structures that begin to appear between 10-20$^{\circ}$ in latitude and can persist for 10s of Myrs (see \cite{sands2025fermibubblesagngammaray} for discussion of these features). In general, more stringent cuts to emission along the line-of-sight result in more variation in the residuals. We compare to the Fermi-LAT galactic center excess within a 10$^{\circ}$ ROI shown in \cite{2017ApJ...840...43A}; we include the high-energy tail of this excess in order to consider the possibility that it may reflect time-dependent phenomena from CRs. 

Galaxy \textbf{m12f} displays more variation in the residual spectra at high energies relative to galaxy \textbf{m12m}, which is more variable at lower ( $<$ 10 GeV) energies. These behaviors reflect differences in the morphology of the two galaxies: \textbf{m12m} has a more distinct galactic bar and bulge, as well as a burstier star formation history, which result in greater variation in the gas density (and thus $\Pi^0$ decay) over time. \textbf{m12f} has a more uniform star formation history, and changes to the high-energy tail of the $\gamma$-ray spectrum are driven primarily by changing radiation and magnetic field strength over time, which preferentially affect losses due to synchrotron and inverse Compton scattering \cite{Sands2025}. 

\begin{figure*}
    \centering
    \includegraphics[width=\linewidth]{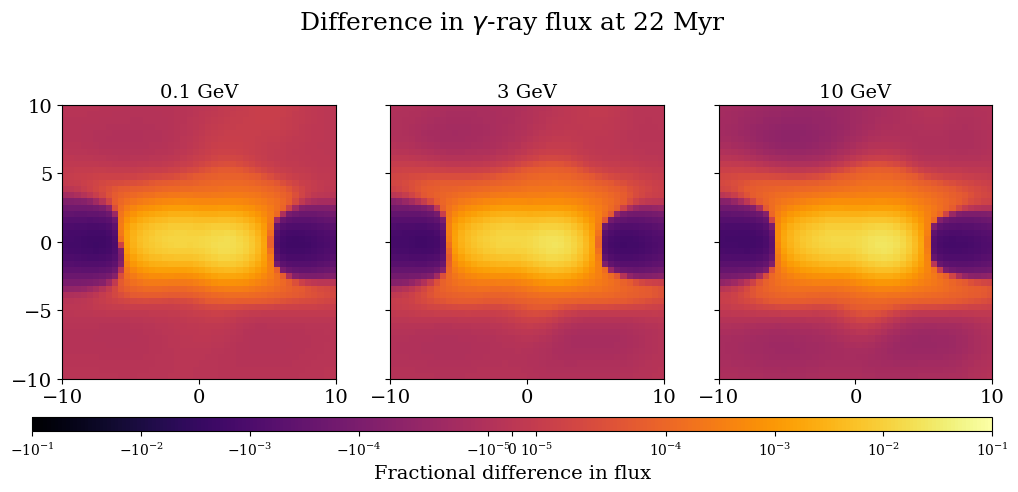}
    \includegraphics[width=\linewidth]{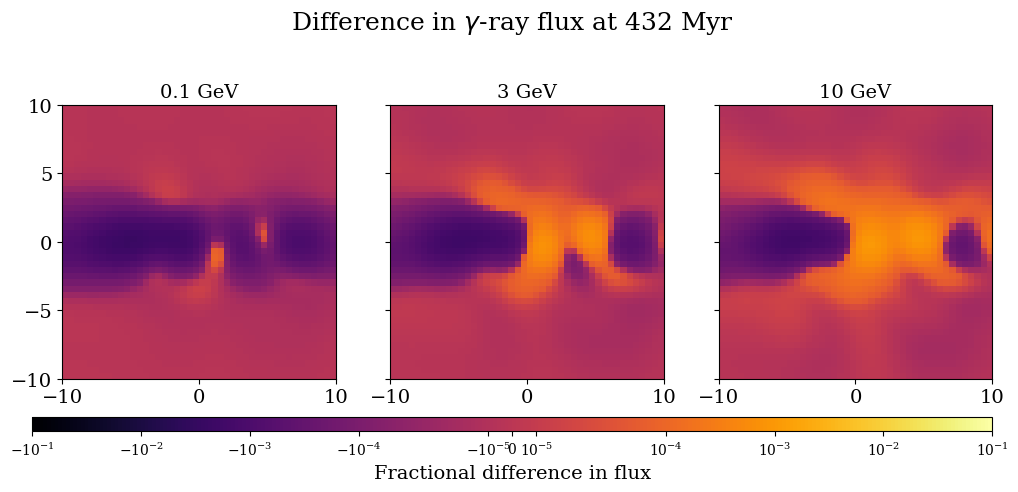}
    \caption{\textbf{Top}: An example of residual flux for \textbf{m12m} in which a large, central starburst creates similar excesses in the 0.1, 3, and 10 GeV channels. \textbf{Bottom}: An example of residual flux for \textbf{m12m} at an earlier snapshot in time, in which a pair of smaller starbursts produce an excess in the 3 and 10 GeV channels that is absent in the 0.1 GeV channel. In both cases, the excesses are surrounded by deficits in flux relative to the template, due to the supernovae vacating the gas in their surrounding regions.}
    \label{fig:morphologies}
\end{figure*}

The morphology of the residual $\gamma$-ray flux across all snapshots most commonly results in similar excesses and deficits across the 0.1 GeV, 3 GeV, and 10 GeV energy channels (Figure \ref{fig:morphologies}, top). There are also snapshots that feature an excess in one or two channels that is not present in the third (Figure \ref{fig:morphologies}, bottom). The excesses can vary from 1 degree to greater than 10 degrees in diameter, and can occur anywhere within the galactic disk (i.e. wherever a supernova or cluster of supernovae has recently occurred). In the following subsections, we consider several examples that illustrate classes of behaviors. 

\subsection{Example 1: Large starburst}

We first consider a snapshot in galaxy \textbf{m12m}, in which there is a significant uptick in star formation within the central kpc in the last 50 Myr of the simulation, leading to a large starburst and an increase in $\gamma$-ray flux. The residual flux in this snapshot is centered, and extends for approximately 8 degrees in latitude and 12 degrees in longitude. The residuals are nearly identical in shape and magnitude across the 0.1 GeV, 3 GeV, and 10 GeV channels. 

Notably, a deficit in $\gamma$-ray flux relative to the template occurs within the plane of the galactic disk, on either side of the central excess. This feature results from the starburst pushing gas away from the galactic center, vacating the surrounding region. These deficits are a common feature across the range of residual maps produced in this analysis, and nearly always appear next to or surrounding an excess in flux. 

\subsection{Example 2: Excess present in higher energy channels and absent from low energy channel}

In this snapshot, which occurs earlier in the evolution of galaxy \textbf{m12m}, there is an excess in flux present in the 3 GeV and 10 GeV channels, which does not appear in the 0.1 GeV channel. This residual pattern is the result of two events that occurred within proximity of each other in space and time, one centered at 0$^{\circ}$ in longitude, and the other at  5$^{\circ}$. This configuration creates an environment in which CR loss rates are more rapid for $\Pi^0$ decay and relativistic Bremsstrahlung (which both depend on gas density) than for ICS, which contributes approximately equally in all three channels \cite{Sands2025}. 

\subsection{Example 3: Centered excess with similarities to GCE}

\label{sec:GCE_ex}

\begin{figure}
    \centering
    \includegraphics[width=\columnwidth]{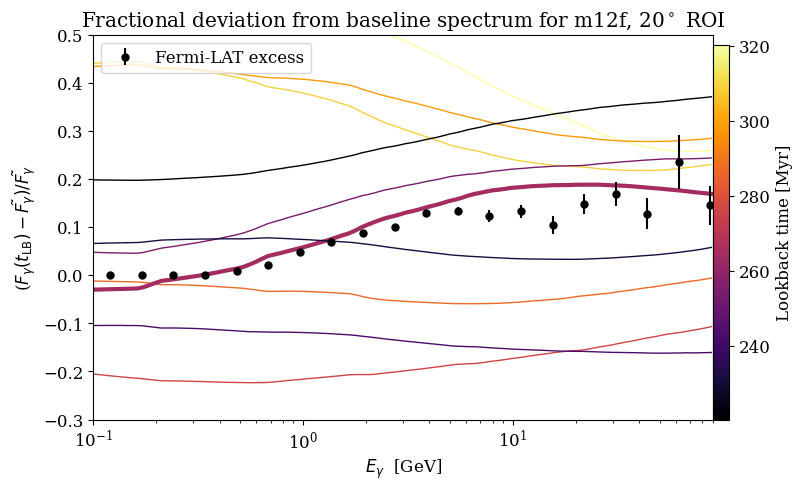}
    \caption{Residual spectra for galaxy \textbf{m12f} for a subset of samples over an 100 Myr timespan. The measured Fermi-LAT $\gamma$-ray excess is shown by the black dotted line. One spectrum similar to the Fermi-LAT excess is indicated by a thicker line; this is the example explored in Section \ref{sec:GCE_ex}, and depicted in Figure \ref{fig:GCE_like}. This set of residuals show a wide variety of shapes and amplitudes, indicating that ``excesses" can occur at any range of energies between 0.1 and 100 GeV.}
    \label{fig:residual_spectra}
\end{figure}

\begin{figure*}
    \centering
    \includegraphics[width=\linewidth]{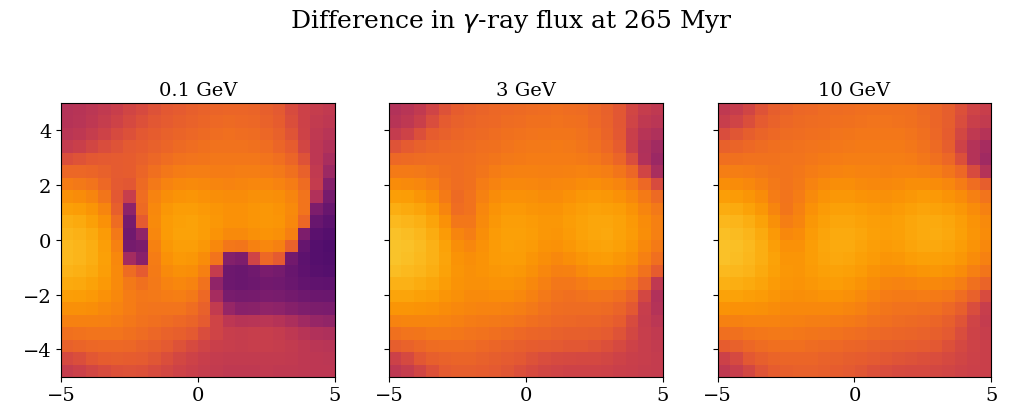}
    \includegraphics[width=\linewidth]{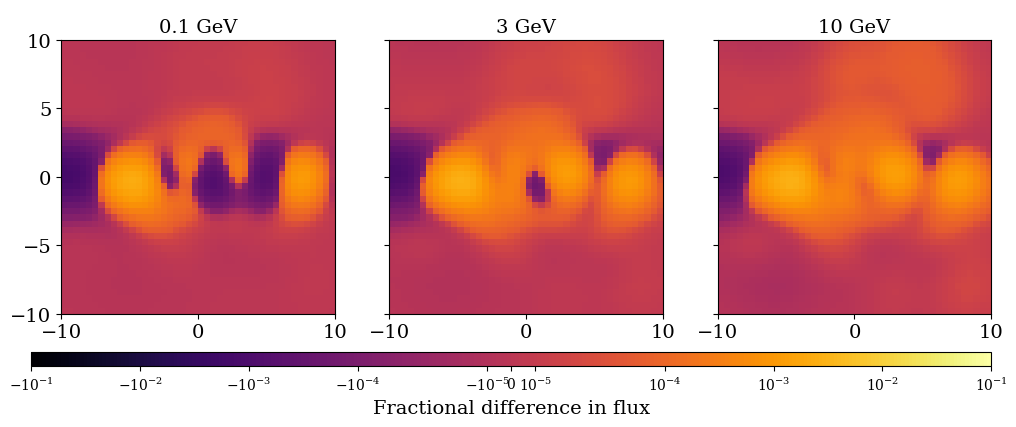}
    \caption{An example of residual flux for \textbf{m12f}, for a snapshot in which the residual spectrum resembles the measured Fermi-LAT galactic center $\gamma$-ray excess; the 10$^{\circ}$ ROI is shown on top, and the 20$^{\circ}$ ROI on the bottom. The morphology is approximately spherical within the central 10$^{\circ}$, with some substructure occurring where there were several recent clustered supernovae.}
    \label{fig:GCE_like}
\end{figure*}

This snapshot of galaxy \textbf{m12f} corresponds to a residual spectrum that is comparable in amplitude and shape to the measured Fermi-LAT excess (Figure \ref{fig:residual_spectra}). The morphology of the residual flux is approximately spherical within the central 10$^{\circ}$ in the 3 GeV and 10 GeV channels,  with some of the excess extending out to 20$^{\circ}$ (Figure \ref{fig:GCE_like}). However, within the 20$^{\circ}$ ROI, deficits occur beside the excesses within the plane of the galactic disk, as observed in the two prior examples. This snapshot displays the most dynamical environment and complexity of the illustrative examples considered in this paper: the $\gamma$-ray flux is the product of at least three clusters of supernovae occurring separately within the disk. This configuration is comparable to the multiple leptonic outburst scenario explored in \cite{Cholis2015}; however, here, the outbursts are still predominantly hadronic, with differences in CR loss times driven by the properties of the multi-phase ISM, with spatial and temporal inhomogeneities in gas density, magnetic and radiation field strength, cosmic ray flux, and star formation rates. While galaxy \textbf{m12f} is not a perfect Milky Way analog, we present this snapshot as an example of a scenario in which a complex, highly dynamical astrophysical environment can produce a residual flux that is phenomenologically similar to the observed Fermi-LAT galactic center $\gamma$-ray excess, simply from non-equilibrium CR transport.

\section{Discussion and Implications for the observed galactic center $\gamma$-ray excess}

The purpose of this exercise is to determine the extent to which time-dependent cosmic-ray transport can affect the interpretation of diffuse $\gamma$-ray emission within the galactic center. Indeed, by comparing the diffuse emission at each snapshot to a baseline template approximating steady-state equilibrium, we find transient bumps and wiggles in the residual $\gamma$-ray spectra occurring at all energies between 0.1 and 100 GeV. We show these residuals for a subset of 10 snapshots for galaxy \textbf{m12f} in Figure \ref{fig:residual_spectra}; the full range of residuals are even more variable in shape and amplitude. Within just this 100 Myr span, there are both deficits and excesses, some significantly larger than the measured Fermi-LAT excess, without a global, monotonic trend. 

These residual excesses are driven by both star formation and its effects on the surrounding environment within the galactic center, and time-dependent CR transport. Within the central kpc, the CR diffusion time is approximately $\tau_{\rm{diff}} \sim (1 \rm{\, kpc})^2/ 3 \times 10^{29} \rm {cm^2 \, s^{-1}} \sim$ 1 Myr, which is comparable to the time scale on which the ISM and the GC $\gamma$-ray spectrum change. Thus, there is typically not sufficient time for CRs injected by supernovae to diffuse throughout the galactic center and reach a globally uniform distribution before the intervening medium is disrupted. Moreover, the efficiencies of the relevant CR loss mechanisms scale differently with CR energy, and have different dependencies on the properties of the local ISM; for instance, inverse Compton and synchrotron loss times are shorter than the $\Pi^0$ decay loss time at 10 GeV \cite{Hopkins_CR, Sands2025}. The differing energy and density dependencies of these loss channels mean that local inhomogeneities in the ISM produce spatially and spectrally variable $\gamma$-ray emission that cannot be captured by smooth, averaged templates.

Moreover, the two simulations considered in this work use our favored model for CR transport as found in \cite{Hopkins_CR} (i.e., the CR injection spectrum and scattering rate are both power-law functions of momentum/rigidity, with $j(R) \propto R^{-4.2}$ and $\bar{\nu} \propto \bar{\nu}_0 R^{-0.5}$, where $\bar{\nu}_0 \approx 10^{-9}\ \rm{s}^{-1}$). While this model has been shown to reproduce constraints on CR transport from local ISM observations, including proton and electron spectra, the boron-to-carbon ratio, and the electron-to-proton ratio, it is far from the only viable model of CR transport \cite{Hopkins2025}. Indeed, other analytical and numerical studies have considered a wide range of physically-motivated models of CR transport that depend on plasma properties \cite{Weber2025, Thomas2025, 2021ApJ...922...11A, 2025ApJ...988..269B, 2021ApJ...913..106B}. These models, which may have non-uniform CR injection spectra or dependence on magnetic field substructure, are likely to produce more small-scale variation than is present in our simulations; thus, our work likely underestimates variation in the galactic center due to small-scale inhomogeneities in the plasma.

Despite these limitations, the results of this exercise have important implications for the interpretation of the \textit{Fermi} GCE. The most conservative interpretation of this exercise is that, by using analysis tools that implicitly assume steady-state equilibrium for cosmic ray propagation, past analyses of the Fermi-LAT observations of the Milky Way's galactic center have significantly underestimated astrophysical uncertainties arising from completely mundane sources. A more radical interpretation of this exercise is that an uptick in diffuse emission can explain the observed galactic center $\gamma$-ray excess without invoking either annihilating or decaying dark matter, or a novel population of pulsars. In this scenario, the Fermi-LAT galactic center $\gamma$-ray excess would be the result of the ``look-elsewhere effect". The Milky Way's galactic center is the only such region of any galaxy that the Fermi-LAT can observe in detail, and it could be merely a coincidence that, at present, the observed spectrum contains a 1-10 GeV excess. In either case, it is imperative to re-evaluate the templates used in analyses of \textit{Fermi}-LAT data, and, if necessary, devise new templates and methods that can fully capture the effects of time-dependent CR transport as well as a wider range of astrophysical configurations. Both GALPROP and empirical template-fitting approaches (the latter of which are often calibrated to GALPROP) share similar limitations in this regard; both assume smooth distributions of stars, gas, magnetic field structure, and radiation fields, and, as static solvers, exclude time dependence by construction \cite{2015JCAP...03..038C, 2016PDU....12....1D, 2017ApJ...840...43A}. Thus, neither of these standard approaches is capable of capturing the time-dynamical effects discussed in this work. 

A limitation of this exercise is that it is based on a sample of two simulations that, while analogous to the Milky Way in many aspects, are not identical to it. As discussed in Section \ref{sec:methods}, of particular concern is the star formation rate in the central kpc of these simulations, which is, throughout the period of the simulation considered, 1-2 orders of magnitude higher than the star formation rate in the Milky Way's Central Molecular Zone (CMZ). The higher star formation rate correlates to a higher supernova rate, and thus the ROI within these simulations may be more violent than the same region within the Milky Way. Previous studies have demonstrated that including a central black hole similar to Sagittarius A* and various channels of AGN feedback, including CR-dominated jets, can suppress star formation in L* galaxies \cite{Mercedes_Feliz_2023, Wellons_2023, 2024MNRAS.532.2724S, 2026MNRAS.545f2060S}. While including a central AGN in these simulations would likely lower the star formation rate, the injection of CRs by AGN jets would likely also contribute to small-scale and temporal variability in diffuse $\gamma$-ray emission. Moreover, with only two simulated galaxies, we cannot robustly quantify the frequency with which mundane CR physics produces residuals resembling the Fermi GCE, which would be necessary to make a rigorous statistical statement about the look-elsewhere effect. Ultimately, we intend for this work to serve as a relatively simple proof-of-concept that can be replicated on other galaxy simulations with varying CR and AGN physics, as well as different star formation histories and galactic center geometries, in order to determine how sensitive these time-dependent effects are to particular choices of CR transport models.

If other galactic centers are found to have excesses at other ranges in the $\gamma$-ray spectrum (including in the MeV and TeV ranges), this evidence would strongly bolster a time-dependent astrophysical explanation; if other galactic centers all contained a 1-10 GeV excess, then such observations would make a compelling argument in favor of the dark matter explanation. While such data does not exist at present, future observations from the Cherenkov Telescope Array Observatory (CTAO), the Compton Spectrometer and Imager (COSI), or a next-generation space-based $\gamma$-ray telescope may be able to shed light on these various hypotheses. In light of the results of this study, we argue that time-dependent CR transport should be seriously considered as a source of at least part of the observed Fermi-LAT galactic center $\gamma$-ray excess. A future exercise that would be highly informative in testing this scenario would be to apply the templates and CR-propagation models used to analyze the \textit{Fermi} GCE to these simulations. Because we know the truth values of CR and $\gamma$-ray flux, gas column density, magnetic fields, as well as the dark matter density and astrophysical J-factor in the simulation, we would be able to compare the best-fit model and identify specific failure modes.

\section*{Acknowledgments}

We thank Tracy Slatyer, Dan Hooper, Lina Necib, Nicholas Rodd, Benjamin Safdi, Rebecca Leane, Joshua Foster, Carlos Blanco, Christopher Dessert, and Caleb Gemmell for insightful conversations. All simulations were run and analyzed on allocations AST21010 and AST20016, supported by TACC and NSF. Support for ISS was provided by the MIT School of Science Dean's office. Support for PFH was provided by a Simons Investigator Grant.

\bibliography{apssamp}

\end{document}